# Rapid change of multiplicity fluctuations in system size dependence at SPS energies

**Andrey Seryakov for the NA61/SHINE Collaboration**

Saint Petersburg State University, Russia
andrey.seryakov@cern.ch

**Abstract.** Recent preliminary results on multiplicity fluctuations in p+p, Be+Be and Ar+Sc collisions from the NA61/SHINE collaboration are presented. The scaled variance of charged hadron multiplicity changes little when going from p+p to Be+Be collisions and drops dramatically from Be+Be to Ar+Sc interactions. The centrality selection procedure and the influence of volume fluctuations are discussed. Comparisons with the EPOS event generator are shown.

## 1. Introduction

This paper presents preliminary results on event-by-event multiplicity fluctuations in inelastic p+p and the most central Ar+Sc and Be+Be collisions. The data were recorded by the NA61/SHINE experiment at the CERN Super Proton Synchrotron (SPS) as a part of the energy and system size scan program. This program aims to study the phase transition between hadron gas and quark gluon plasma in heavy ion collisions and to find the critical point. The expected signal of a critical point is a non-monotonic behavior of different fluctuation measures including multiplicity fluctuations [1].

The paper is organized as follows. In sec.2 two measures of multiplicity fluctuations are introduced: the scaled variance and the strongly intensive quantity Ω [2]. The event and centrality selection procedure is discussed in sec.3. The NA61/SHINE acceptance, track selection, uncertainties and corrections are described in sec.4 and sec.5. Results are presented in sec.6 and discussed in sec.7.

## 2. Multiplicity and forward energy fluctuations

The most common way to measure multiplicity fluctuations is to use a scaled variance ω of the distribution:

$$\omega[N] = \frac{\langle N^2\rangle - \langle N\rangle^2}{\langle N\rangle} \quad (1)$$

where N is the multiplicity of charged particles in an event, and $\langle\ \rangle$ denotes the averaging over events. The scaled variance is an intensive quantity which does not depend on the volume of the system within the grand canonical ensemble (GCE) of statistical mechanics or on the number of sources within models of independent sources like the wounded nucleon model (WNM). The scaled variance has the value zero if there are no multiplicity fluctuations and unity for a Poisson multiplicity distribution.

Unfortunately, the volume of the system created in heavy ion collisions cannot be fixed and fluctuates a lot event by event. As the scaled variance depends on the fluctuations of the volume [3] the problem of reducing this effect becomes very important in fluctuation studies. Two different approaches were implemented to minimize the effect:

- constrain fluctuations of the number of participating nucleons by selecting narrow centrality classes. This will be discussed in the next section.
- use strongly intensive quantities in the analysis, which by definition do not depend on volume fluctuations (in GCE and WNM) [2-3].

The current analysis uses the strongly intensive quantity $\Omega[\mathrm{A},\mathrm{B}]$:

$$\Omega[\mathrm{A},\mathrm{B}] = \omega[\mathrm{A}] - \frac{\langle \mathrm{AB}\rangle - \langle \mathrm{A}\rangle\langle \mathrm{B}\rangle}{\langle \mathrm{B}\rangle} \qquad (2)$$

It was shown in Ref. [2] that if A and B are uncorrelated from the fixed volume of the GCE or a single source in the WNM then $\Omega[\mathrm{A},\mathrm{B}]$ is equal to the scaled variance of the quantity A from the fixed volume (GCE) or a single source (WNM). To obey this condition A is selected as the multiplicity of produced particles and B as the projectile participant energy $\mathrm{E_p} = \mathrm{E_{beam}} - \mathrm{E_{PSD}}$, where $\mathrm{E_{beam}}$ is the total beam energy and $\mathrm{E_{PSD}}$ is the energy measured by the PSD, the NA61/SHINE hadron calorimeter positioned exactly on the beam line [4].

## 3. Event selection and centrality determination

The crucial part of any fluctuation analysis is the centrality selection procedure. NA61/SHINE has a unique possibility as a fixed-target experiment to measure the total energy of non-interacting projectile nucleons as there is no need for a beam hole. In order to achieve this the hadron calorimeter PSD was placed on the beam line downstream of the Time Projection Chambers (TPCs). More information about NA61/SHINE detector systems can be found in Ref. [4]. The PSD detector consists of 44 independent modules. Selection of different sets of modules for each colliding system and energy allows to minimize the impact of produced particles on the calorimeter signal and maximize the contribution from non-interacting projectile nucleon spectators.

In order to reduce autocorrelation between multiplicity of charged particles used for fluctuation analysis and signals utilized for the centrality determination in the NA61/SHINE experiment, only the energy deposited in the PSD calorimeter is used for event selection. Event-by-event multiplicity measured by the TPCs does not take part in the procedure.

The exact way of experimental centrality determination has a large effect on the multiplicity fluctuation measurement. Therefore, for a proper comparison between models and data, theoreticians have to repeat the experimental procedures as closely as possible, because quantities obtained with centrality selection based on multiplicity, impact parameter or forward energy may differ a lot, especially for light and intermediate size nuclei such as Be and Ar. The NA61/SHINE centrality selection procedure is fully based on forward energy. To calculate this energy in a Monte-Carlo event generator the PSD acceptance maps [5] should be used. Then the 5% of the events with the lowest energy in the acceptance will correspond to the NA61/SHINE results for 5% centrality selection.

## 4. Track selection and acceptance

Track selection criteria were chosen in a way to select only primary charged hadrons as well as to reduce the contamination from secondary interactions, weak decays and pile-up events.

The multiplicity fluctuations strongly depend on the acceptance in which they are measured. Therefore, it is highly important for any model comparison to use the same acceptance. The NA61/SHINE acceptance maps can be found in Ref. [6]. Moreover, in the current paper only particles with transverse momenta $\mathrm{p_T} < 1.5\ \mathrm{GeV/c}$ and rapidity with pion mass assumption $0 < y_\pi < \mathrm{y_{beam}}$ were analysed.

## 5. Statistical and systematic uncertainties, corrections

Statistical uncertainties were calculated using the sub-sample method. Analysis of systematic uncertainties is not finished yet but they are expected to be smaller than 5%.

Results for Ar+Sc and Be+Be collisions presented below are not corrected for experimental biases. To minimize such biases only results for negatively charged hadrons are shown. Nevertheless, to estimate the magnitudes of detector biases, an analysis of events generated by the EPOS event generator was performed at two levels: before and after detector response simulated with the GEANT based NA61/SHINE program chain. A comparison of the results is shown in Fig. 1. The differences between results are much smaller than 5% except at the top energy where they are on the level of 5%.
The p+p data is fully corrected for detector effects and off-target interactions. The procedure is described in detail in Ref. [7].

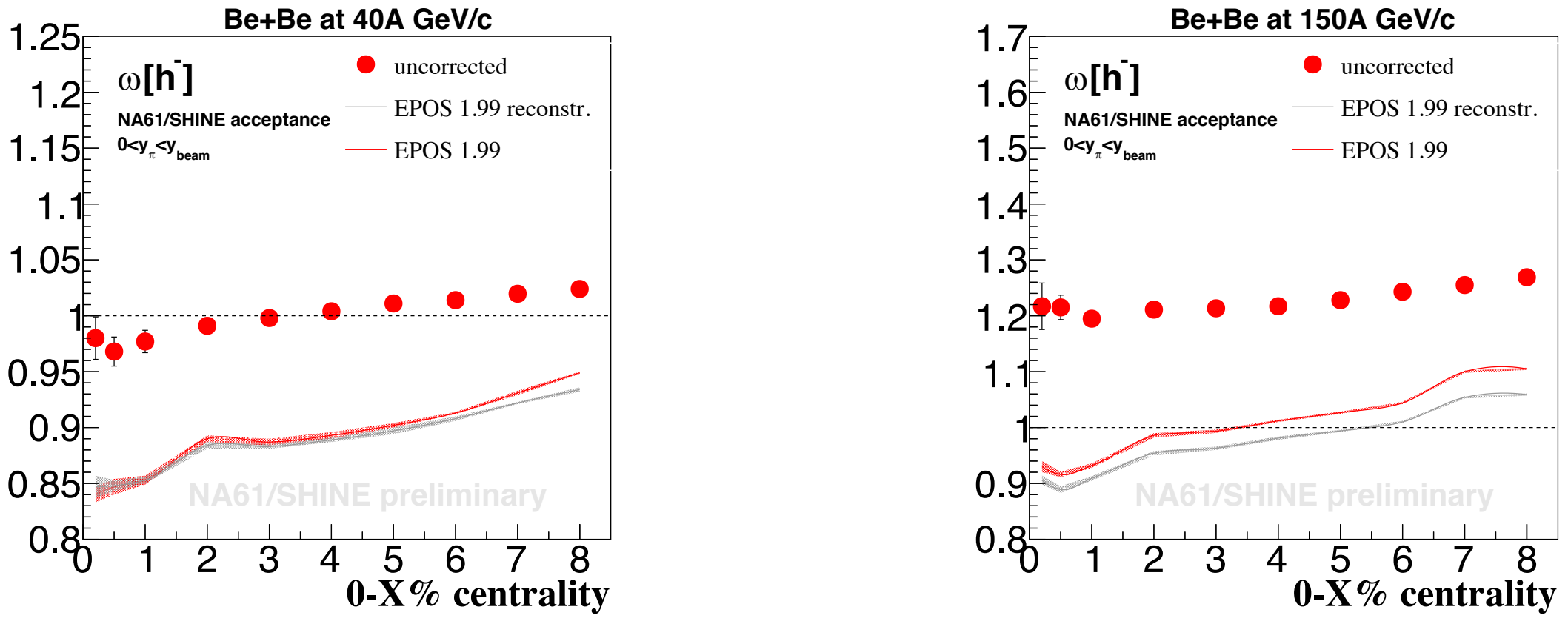

**Figure 1.** Centrality dependence of $\omega[h^-]$ in Be+Be collisions at 40*A* (left) and 150*A* (right) GeV/c. Uncorrected preliminary NA61/SHINE data are shown by red dots, pure EPOS predictions in the NA61/SHINE acceptance by the red lines and reconstructed EPOS after detector simulation by the grey lines. The two presented energies show the largest differences between pure and reconstructed simulation.

## 6. Results

The preliminary results shown in this section refer to accepted primary negatively charged hadrons produced in centrality selected Be+Be and Ar+Sc collisions and in inelastic p+p interactions.
The first observation is that the strongly intensive quantity $\Omega[h^-, E_P]$ does not depend on the width of the centrality interval as expected, contrary to the scaled variance of the multiplicity distribution (see Fig. 2). Both quantities converge to the same common limit for the most central collisions.

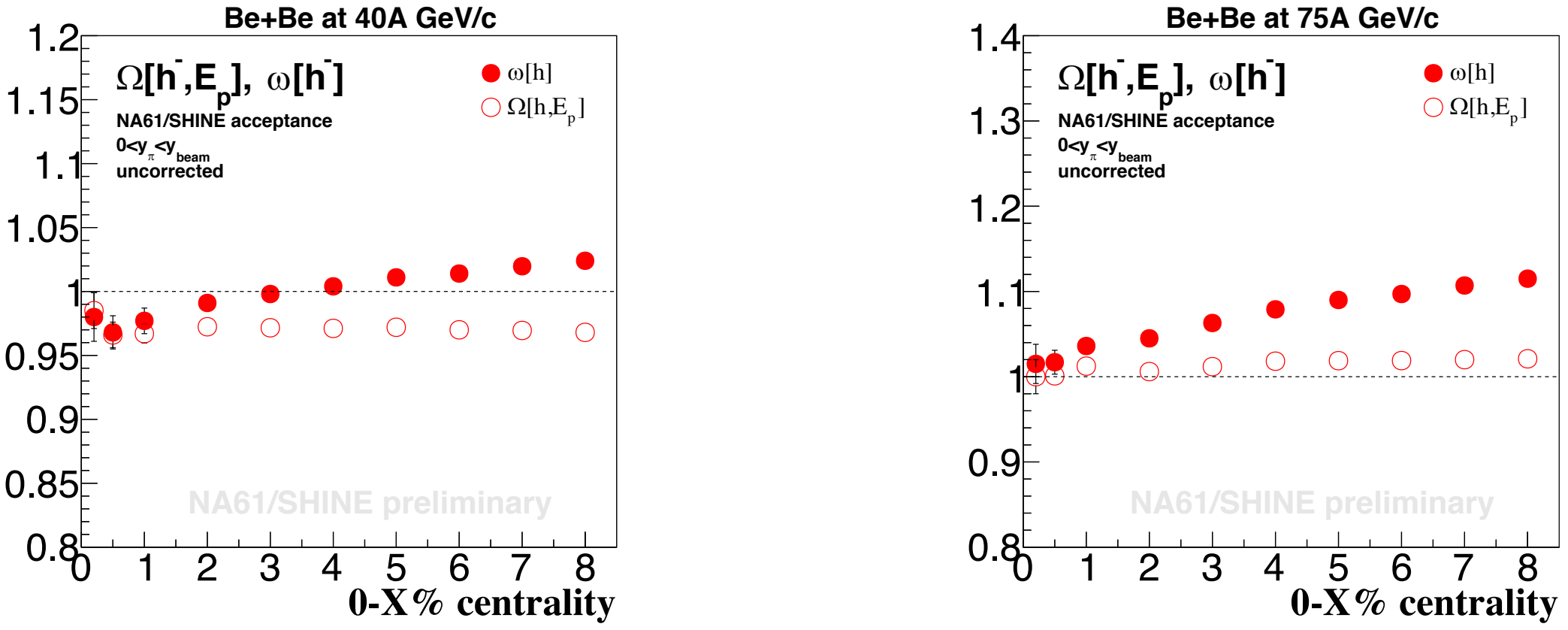

**Figure 2.** Centrality dependence of $\omega[h^-]$ and $\Omega[h^-, E_P]$ in Be+Be collisions at 40*A* (left) and 75*A* (right) GeV/c.

Figure 3 shows the energy dependence of $\omega[h^-]$ and $\Omega[h^-, E_P]$ in central Be+Be and Ar+Sc events, as well as results for $\omega[h^-]$ in p+p collisions. The systems created in p+p and the most central Be+Be collisions show very similar increase for all SPS energies. The energy dependence for Ar+Sc collisions is strikingly different. Fluctuations do not increase with beam energy and are significantly below the p+p and Be+Be data.

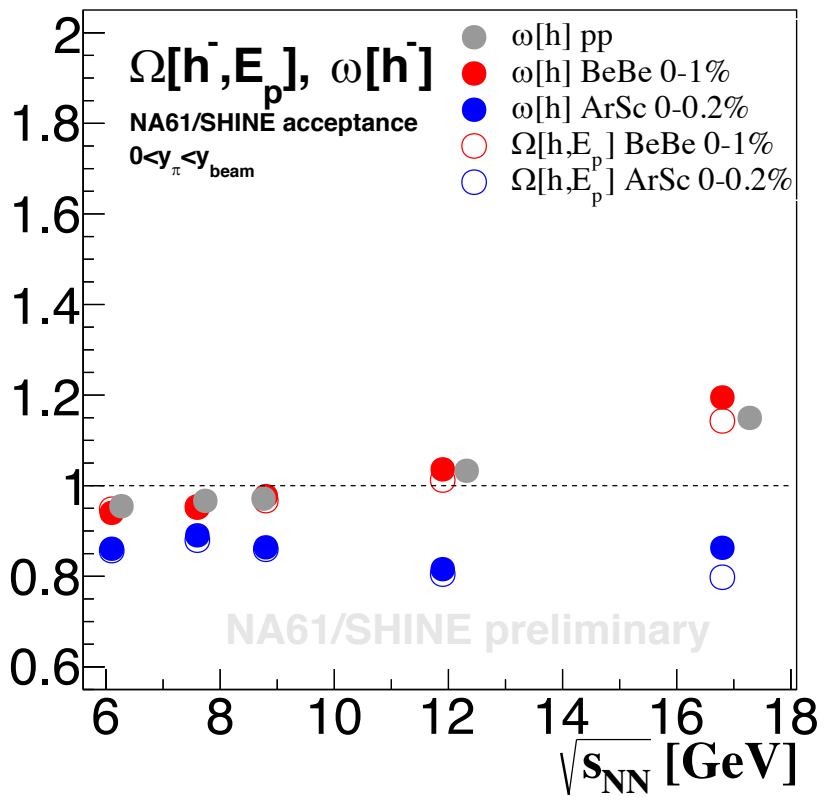

**Figure 3.** Energy dependence of negatively charged hadron multiplicity fluctuations obtained by NA61/SHINE at forward rapidity $0 < y_\pi < y_{beam}$ and transverse momenta $p_T < 1.5$ GeV/c. The scaled variance is shown for inelastic p+p (grey dots), 0-1% Be+Be (full red dots) and 0-0.2% Ar+Sc (full blue dots) collisions and the strongly intensive quantity $\Omega[h^-, E_P]$ by corresponding open dots. Only statistical uncertainties are shown.

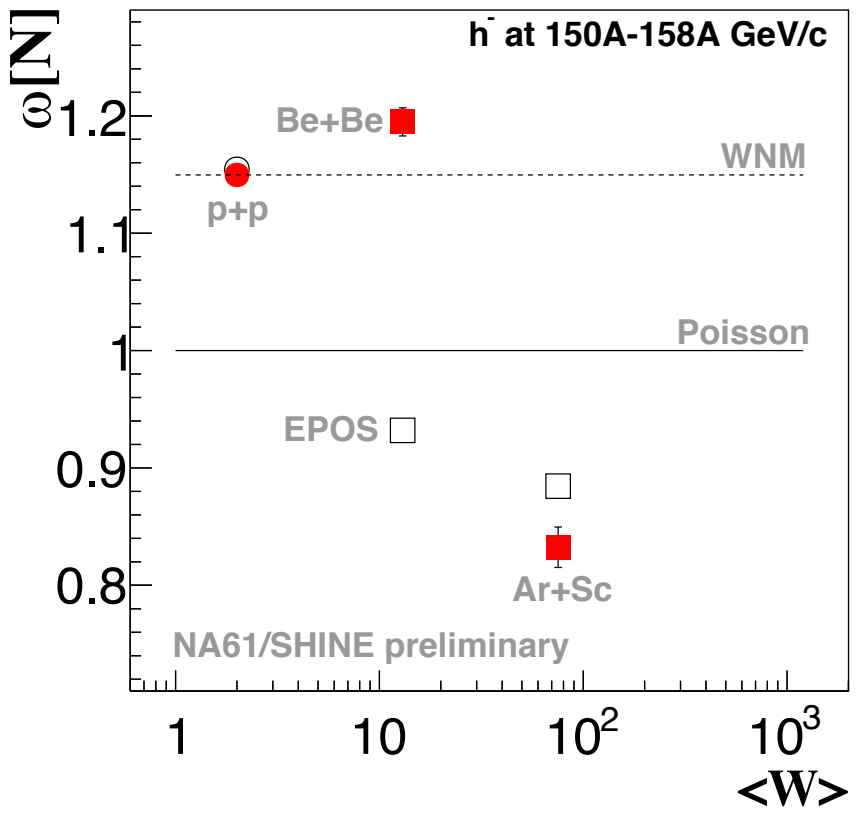

**Figure 4.** Scaled variance of the multiplicity distribution of negatively charged hadrons versus mean number of wounded nucleons at high SPS energy. Results for inelastic p+p interactions as well as central 0-1% Be+Be and 0-0.2% Ar+Sc collisions are shown at forward-rapidity, $0 < y_\pi < y_{beam}$, and in $p_T < 1.5$ GeV/c. Only statistical uncertainties are shown. The NA61/SHINE data are compared to the EPOS1.99 model predictions (open squares).

## 7. Conclusions

New measurements of multiplicity fluctuations were performed by the NA61/SHINE experiment at CERN SPS energies using the intensive quantity ω and the strongly intensive quantity $\Omega[h^-, E_P]$. These two measures converge to common limit for the most central collisions as expected. While ω increases with the size of the centrality interval, $\Omega[h^-, E_P]$ does not change. Thus the strongly intensive quantity Ω better characterizes features of individual sources in heavy ion collisions.

The suppression of multiplicity fluctuations in the most central Pb+Pb collisions compared to inelastic p+p reactions was previously noted in Ref. [7]. New results from Ar+Sc collisions confirm this effect for intermediate mass nuclei: the Ar+Sc result is much below that from p+p collisions.

Unexpectedly, the light nuclei system (Be+Be) shows behavior very similar to that of p+p, while multiplicity fluctuations in Ar+Sc are suppressed. If this effect is due to different properties of particle sources in p+p and Be+Be versus Ar+Sc, it may be an indication of some type of transition in heavy ion collisions, namely, rapid change of hadron production properties when moving from the Be+Be to the Ar+Sc colliding system. The effect can be interpreted as the beginning of creation of large clusters of strongly interacting matter and may be explained by percolation models [8-12] or AdS/CFT correspondence [13]. This experimental phenomenon was referred to as the *onset of fireball* (by Edward Shuryak at the CPOD 2017 conference).

## Acknowledgements

This work was supported by the Russian Science Foundation under grant 17-72-20045.